\documentclass[prc,aps,twocolumn,showpacs]{revtex4}
\usepackage{amsmath,amssymb}
\usepackage{graphicx}
\def\pcomma{$^,$}
\def\pmainz{$^1$}
\def\plpi{$^{2}$}
\def\ptomsk{$^{3}$}
\def\pucla{$^4$}
\def\pglsg{$^5$}
\def\pkent{$^6$}
\def\pbonn{$^7$}
\def\pgatch{$^8$}
\def\pgiess{$^9$}
\def\ppavia{$^{10}$}
\def\pgwu{$^{11}$}
\def\pedinb{$^{12}$}
\def\psackv{$^{13}$}
\def\plund{$^{14}$}
\def\pbasel{$^{15}$}
\def\pinr{$^{16}$}
\def\pzagreb{$^{17}$}
\def\pcua{$^{18}$}
\begin{document}

\title{\boldmath
Study of the $\gamma p\to \pi^0\pi^0 p$ reaction with the Crystal Ball/TAPS at the Mainz
Microtron}

\author{
V.~L.~Kashevarov\pmainz\pcomma\plpi\footnote[1]{Electronic address: kashev@kph.uni-mainz.de},
A.~Fix\ptomsk\footnote[2]{Electronic address: fix@mph.phtd.tpu.ru},
S.~Prakhov\pucla,
P.~Aguar-Bartolom\'e\pmainz,
J.~R.~M.~Annand\pglsg,
H.~J.~Arends\pmainz,
K.~Bantawa\pkent,
R.~Beck\pbonn,
V.~Bekrenev\pgatch,
H.~Bergh\"auser\pgiess,
A.~Braghieri\ppavia,
W.~J.~Briscoe\pgwu,
J.~Brudvik\pucla,
S.~Cherepnya\plpi,
R.~F.~B.~Codling\pglsg,
E.~J.~Downie\pmainz\pcomma\pglsg,
P.~Drexler\pgiess,
L.~V.~Fil'kov\plpi,
D.~I.~Glazier\pedinb,
R.~Gregor\pgiess,
E.~Heid\pmainz\pcomma\pgwu,
D.~Hornidge\psackv,
L.~Isaksson\plund,
I.~Jaegle\pbasel,
O.~Jahn\pmainz,
T.~C.~Jude\pedinb,
I.~Keshelashvili\pbasel,
R.~Kondratiev\pinr,
M.~Korolija\pzagreb,
M.~Kotulla\pgiess,
A.~Koulbardis\pgatch,
S.~Kruglov\pgatch,
B.~Krusche\pbasel,
V.~Lisin\pinr,
K.~Livingston\pglsg,
I.~J.~D.~MacGregor\pglsg,
Y.~Maghrbi\pbasel,
D.~M.~Manley\pkent,
J.~C.~McGeorge\pglsg,
E.~F.~McNicoll\pglsg,
D.~Mekterovic\pzagreb,
V.~Metag\pgiess,
A.~Mushkarenkov\ppavia,
B.~M.~K.~Nefkens\pucla,
A.~Nikolaev\pbonn,
R.~Novotny\pgiess,
H.~Ortega\pmainz,
M.~Ostrick\pmainz,
P.~Ott\pmainz,
P.~B.~Otte\pmainz,
B.~Oussena\pmainz,
P.~Pedroni\ppavia,
F.~Pheron\pbasel,
A.~Polonski\pinr,
J.~Robinson\pglsg,
G.~Rosner\pglsg,
T.~Rostomyan\ppavia\footnote[3]{Present address:
 Institut f\"ur Physik, University of Basel, Switzerland},
S.~Schumann\pmainz\pcomma\pbonn,
M.~H.~Sikora\pedinb,
D.~I.~Sober\pcua,
A.~Starostin\pucla,
I.~I.~Strakovsky\pgwu,
I.~M.~Suarez\pucla,
I.~Supek\pzagreb,
C.~M.~Tarbert\pedinb,
M.~Thiel\pgiess,
A.~Thomas\pmainz,
M.~Unverzagt\pmainz\pcomma\pbonn,
D.~P.~Watts\pedinb,
D.~Werthm\"uller\pbasel,
and F.~Zehr\pbasel
 \\
\vspace*{0.1in}
(Crystal Ball at MAMI, TAPS, and A2 Collaborations)
\vspace*{0.1in}
}

\affiliation{
\pmainz Institut f\"ur Kernphysik, Johannes Gutenberg-Universit\"at Mainz,
D-55099 Mainz, Germany}
\affiliation{
\plpi Lebedev Physical Institute, 119991 Moscow, Russia}
\affiliation{
\ptomsk Laboratory of Mathematical Physics, Tomsk Polytechnic University, Tomsk, Russia}
\affiliation{
\pucla University of California Los Angeles, Los Angeles, California 90095-1547, USA}
\affiliation{
\pglsg Department of Physics and Astronomy, University of Glasgow, Glasgow G12 8QQ, 
United Kingdom}
\affiliation{
\pkent Kent State University, Kent, Ohio 44242-0001, USA}
\affiliation{
\pbonn Helmholtz-Institut f\"ur Strahlen- und Kernphysik, University of Bonn, 
D-53115 Bonn, Germany}
\affiliation{
\pgatch Petersburg Nuclear Physics Institute, 188350 Gatchina, Russia}
\affiliation{
\pgiess II Physikalisches Institut, University of Giessen, D-35392 Giessen, Germany}
\affiliation{
\ppavia INFN Sesione di Pavia, I-27100 Pavia, Italy}
\affiliation{
\pgwu The George Washington University, Washington, DC 20052-0001, USA}
\affiliation{
\pedinb School of Physics, University of Edinburgh, Edinburgh EH9 3JZ, United Kingdom}
\affiliation{
\psackv Mount Allison University, Sackville, New Brunswick E4L 1E6, Canada}
\affiliation{
\plund Lund University, SE-22100 Lund, Sweden}
\affiliation{
\pinr Institute for Nuclear Research, 125047 Moscow, Russia}
\affiliation{
\pzagreb Rudjer Boskovic Institute, HR-10000 Zagreb, Croatia}
\affiliation{
\pbasel Institut f\"ur Physik, University of Basel, CH-4056 Basel, Switzerland}
\affiliation{
\pcua The Catholic University of America, Washington, DC 20064, USA}

\date{today}

\begin{abstract}
The $\gamma p\to \pi^0\pi^0 p$ reaction has been measured from threshold to
1.4 GeV using the Crystal Ball and TAPS photon spectrometers together with the photon
tagging facility at the Mainz Microtron.
The experimental results include total and differential cross sections as well as
specific angular distributions, which were used to extract partial-wave amplitudes.
In particular, the energy region below the $D_{13}(1520)$ resonance was studied.
\end{abstract}


\maketitle

\section{Introduction}
Although studied for a long time, the properties of many baryon resonances are still not
well known and a clear understanding of resonances in QCD is still not possible.
Some states below 2.5 GeV are believed to couple strongly to final states with two
pseudoscalar mesons. Therefore, the investigation of $\pi\pi$ and $\pi\eta$ photoproduction
provides important new information about the nucleon excitation spectrum.

During the last two decades, an extensive study of double-pion photoproduction for
$E_\gamma<1$~GeV has been undertaken
\cite{Bragh,Hart,Wolf,Langgrt,Ahren1,Assa,Kot,Ahren2,Strauch,Ahren3,Thom,Sara,Kram,Mess}.
The theoretical interpretation of the data in various isospin channels was carried out
using different phenomenological analyses~\cite{Oset,Laget,Ochi,Mokeev,FA,Kammano}.
As a rule, the models for double-pion production are based on isobar models or effective
field theories. Typically, the reaction amplitude is constructed as a sum of background and
resonance contributions. The background part contains nucleon Born terms as well as meson
exchange in the $t$ channel. The resonance part is a coherent sum of $s$-channel
resonances decaying into $\pi\pi N$ via intermediate formation of meson-nucleon and
meson-meson states (``isobars''). Despite significant qualitative differences
between the models, in general they provide an acceptable description of the existing
cross-section data. Such an apparent consistency between theoretical models does not 
indicate a high level in understanding double-pion photoproduction; rather it demonstrates 
a weak sensitivity of the existing data to the underlying dynamics.

The reaction $\gamma p\to \pi^0\pi^0 p$ is a typical example where the data on double-pion
production have not been fully understood theoretically, especially below the second
resonance region. A widely accepted property of this reaction is
a large contribution from $D_{13}(1520)$, which is known to couple strongly to the
$\pi\Delta$ channel \cite{Manley,Vrana}. The $D_{13}(1520)$ contribution to
the $\gamma p\to \pi^0\pi^0 p$ total cross section is seen as the first peak
at $E_\gamma \approx 730$~MeV (see Fig.\,\ref{fig1}), the features of which are
reproduced more or less successfully by all models.
However, the dynamics underlying this reaction in the region
from the $D_{13}(1520)$ resonance down to threshold have not been well understood so far.
In this region, the total cross section demonstrates an almost linear rise,
hinting at $s$-wave dominance in the final state, which, however, is not confirmed by
the theory. An attempt to describe such behavior by a large contribution
of the Roper resonance $P_{11}(1440)$, decaying into $\sigma N$ in $s$ wave~\cite{Laget},
seems to be ruled out by subsequent investigations~\cite{Sara,Oset,Ochi,FA,Kammano}.
The $D_{13}(1520)$ contribution itself, according to the results of Refs.\cite{Sara,Ochi,FA},
reduces rapidly with decreasing energy and cannot explain the experimental data
in the region below $E_\gamma=650$~MeV. In Ref.\,\cite{Sara},
the authors try to describe the $\gamma p\to \pi^0\pi^0 p$ reaction by a dominant 
contribution from the $D_{33}(1700)$ resonance.
The well-known minimum at $W=1.6$~GeV and the second maximum at $W=1.7$~GeV, seen
in the $\gamma p\to \pi^0\pi^0 p$ total cross section, were described
in Ref.\,\cite{Sara} by the interference between $D_{13}(1520)$ and $D_{33}(1700)$.
However, a simple consideration using the Clebsch-Gordan coefficients
shows that $\pi\Delta$ photoproduction in the $I=3/2$ channel should lead
to the ratio
\begin{eqnarray}
&&\frac{\sigma(\pi^+\pi^-p)}{\sigma(\pi^0\pi^0p)}\approx
\frac{\sigma(\pi^-\Delta^{++})+\sigma(\pi^+\Delta^{0})}
{\sigma(\pi^0\Delta^+)}\nonumber\\
&&=\Big\{\left(C_{1-1\frac32\frac32}^{\frac32\frac12}C_{11\frac12\frac12}^{\frac32\frac32}
\right)^2
+\left(C_{11\frac32-\frac12}^{\frac32\frac12}C_{1-1\frac12\frac12}^{\frac32-\frac12}
\right)^2\Big\}\nonumber\\
&&\phantom{xx}:\left(C_{10\frac32\frac12}^{\frac32\frac12}C_{10\frac12\frac12}^{\frac32\frac12}
\right)^2=13\,,
\end{eqnarray}
where for simplicity we neglect the interference between two possible
$\pi N$ pairs. Taking a value $\sigma=4$\,$\mu$b for the $D_{33}(1700)$
contribution to $\pi^0\pi^0p$, as predicted in \cite{Sara}, assumes that
at least 70\,$\%$ of the $\gamma p\to\pi^+\pi^-p$ total cross section
comes from this resonance alone, which seems unlikely.
The major (about 80$\%$) part of the $\gamma p\to\pi^+\pi^-p$ total cross
section comes from the $\Delta$-Kroll-Ruderman term, so that the addition of such a strong
contribution from $D_{33}(1700)$ would result in a significant overestimation of
the experimental data.
Thus the dynamics of double-$\pi^0$ photoproduction in the energy region below
$D_{13}(1520)$ are still far from being well understood.

Major disagreements between the results of different models as well as between the
theoretical predictions and the experimental data were revealed in Refs.\,\cite{Strauch} and
\cite{Kram}, in which the measurement of the beam helicity asymmetry
$I^\odot$ for $\pi^+\pi^-$, $\pi^+\pi^0$, and $\pi^0\pi^0$ was reported.
As discussed in Ref.\,\cite{Roca}, this quantity is very
sensitive to the model details, so that even a small variation of the model parameters
can change the results significantly.
However, the interpretation of polarization measurements
in terms of the spin and parity of $J^P$ is quite difficult, especially for the processes
with more than two particles in the final state. Therefore, it is desirable to find a method
that, on the one hand, will be sensitive to the details of the dynamical structure and, on the
other hand, will provide a clear interpretation of the results in terms of spin-parity of the
contributing waves. Furthermore, especially important is that the method should not be
connected strictly to the isobar model, allowing one to perform the partial-wave
analysis with a minimal model dependence.
An approach that seems to obey the requirements discussed above was applied for analysis of
inelastic pion-nucleon scattering $\pi N\to\pi\pi N$ (see, for example, \cite{Arnold,Morgan}).
A similar formalism for photoproduction of two pseudoscalars was developed in Ref.\,\cite{FiAr}.
Such approaches require high-statistics data covering the full solid angle.
In the present work, we remeasured double-$\pi^0$ photoproduction off the proton
with an unprecedented accuracy and applied the formalism of Ref.\,\cite{FiAr}
to study its dynamics, having a main goal to learn
which $J^P$ waves dominate in this reaction at the energies below $D_{13}(1520)$.

The experimental data in the present study were obtained at the Mainz tagger
photon facility using an almost $4\pi$ detector
based on the Crystal Ball and TAPS multiphoton spectrometers.

This paper includes a brief description of the experimental setup, data handling,
the model formalism, discussion of the results, and conclusions.

\section{Experimental setup}
\label{sec:Setup}

The reaction $\gamma p\to \pi^0\pi^0 p$ was studied using the Crystal Ball (CB)\,\cite{CB}
as the central spectrometer and TAPS \,\cite{TAPS,TAPS2} as a forward spectrometer. These
detectors were installed in the energy-tagged bremsstrahlung photon beam of
the Mainz Microtron (MAMI)\,\cite{MAMI,MAMIC}. The photon energies were determined
by the Glasgow tagging spectrometer\,\cite{TAGGER2,TAGGER,TAGGER1}.

The CB detector is a sphere consisting of 672 optically insulated NaI(Tl) crystals, shaped as
truncated triangular pyramids, which point toward the center of the sphere. The crystals are
arranged in two hemispheres that cover 93\% of $4\pi$ sr, sitting
outside a central spherical cavity with a radius of 25~cm, which is designed to hold the 
target and inner detectors. In this experiment, TAPS was
arranged in a plane consisting of 384 BaF$_2$ counters of hexagonal cross section. It was
installed 1.5~m downstream of the CB center covering the full azimuthal range for polar angles
from $1^\circ$ to $20^\circ$. More details on the energy and angular resolution of the CB and
TAPS are given in Refs.\,\cite{slopemamic,etamamic}.

The present measurement used 855-MeV and 1508-MeV
electron beams from the upgraded Mainz Microtron, MAMI-C\,\cite{MAMIC}.
The data with the 1508-MeV beam were taken in 2007, and with the 855-MeV beam
in 2008. Bremsstrahlung photons, produced by the 1508-MeV electrons
in a 10-$\mu$m Cu radiator and collimated by a 4-mm-diameter Pb collimator,
were incident on a 5-cm-long liquid hydrogen (lH$_2$) target located
in the center of the CB. The energies of the incident
photons were measured in the range 617 to 1402~MeV by detecting
the post-bremsstrahlung electrons in the Glasgow tagger\,\cite{TAGGER2}.
With the 855-MeV electron beam, bremsstrahlung photons were produced
in a diamond radiator, collimated by a 3-mm-diameter Pb collimator, and
incident on a 10-cm-long lH$_2$ target. In this experiment, the energies
of the incident photons were tagged from 84 to 796~MeV.
The energy resolution of the tagged photons is mostly defined by the
width of the tagger focal plane detectors, and by the electron beam energy.
For a beam energy of 1508~MeV, a typical width of a tagger channel was about 4~MeV,
and about 2~MeV for a beam energy of 855~MeV.
Due to the beam collimation only part of the bremsstrahlung photon flux
reached the lH$_2$ target. In order to evaluate the reaction cross sections,
the probability of bremsstrahlung photons reaching the target (the so-called tagging
efficiency) was measured for each tagger channel.
The typical tagging efficiency in the experiment with the 1508-MeV electron beam
was found to vary between 67\% and 71\%. With the 855-MeV electron beam, the tagging
efficiency varied with photon energy between 30\% and 60\%.

The experimental trigger in the measurement with the 1508-MeV electron beam
required the total-energy deposit in the CB to exceed $\sim 320$~MeV
and the number of so-called hardware clusters in the CB to be larger than two.
With the 855-MeV electron beam, the trigger required the total energy
in the CB to exceed $\sim 100$~MeV, and the number of hardware clusters in the CB and
TAPS together to be larger than 1, with at least one hardware cluster in the CB.

More details on the experimental conditions of the data taking with the 1508-MeV 
electron beam in 2007 are given in Refs.\,\cite{slopemamic,etamamic}.

\section{Data analysis}\label{Data}

The reaction $\gamma p\to \pi^0\pi^0 p$
was identified using events with four photons detected in the calorimeters.
There were two independent analyses made to crosscheck the results.
In the first analysis, the event-selection procedure was similar
to the one that was used to measure the reaction $\gamma p\to \pi^0\eta p$\,\cite{Kashev}.
The second analysis was based on the kinematic-fit technique and was
similar to those published in Refs.\,\cite{slopemamic,etamamic}.
Both analyses are in excellent agreement. Since the kinematic-fit technique typically
yields data with better resolution, the results of the second analysis were used.

\begin{figure}
\begin{center}
\resizebox{0.49\textwidth}{!}{%
\includegraphics{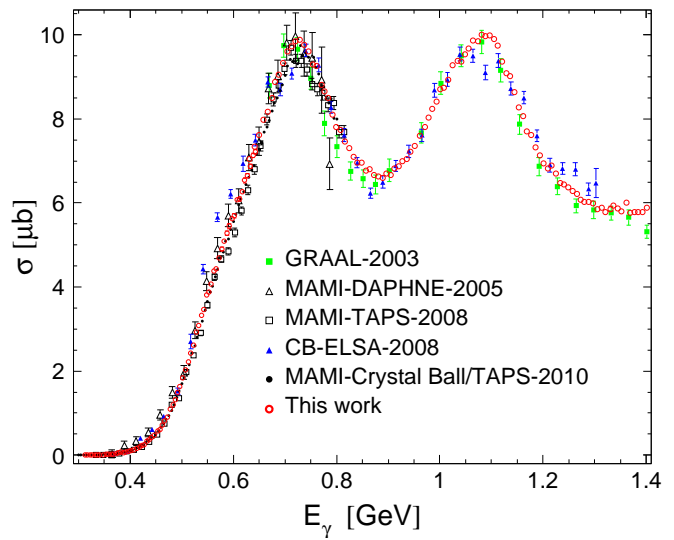}}
\caption{(Color online) Total cross sections for $\gamma p\to\pi^0\pi^0 p$ are shown
 as a function of the incident-photon energy. The results obtained in this work are
 compared to the existing data from  GRAAL\,\cite{Assa}, CB-ELSA\,\cite{Thom,Sara},
 DAPHNE\,\cite{Ahren2}, TAPS\,\cite{Sara}, and Crystal Ball/TAPS\,\cite{Schum}. Only
 statistical uncertainties are shown for all data.}
\label{fig1}
\end{center}
\end{figure}

\begin{figure}
\begin{center}
\resizebox{0.49\textwidth}{!}{%
\includegraphics{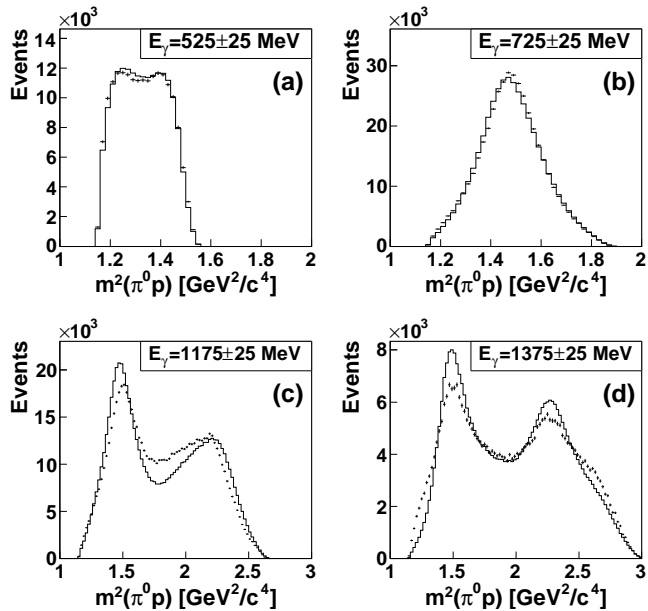}}
\caption{Experimental $m^2(\pi^0p)$ invariant-mass distributions (crosses)
 compared to those obtained from the MC simulation (solid line) of the
 $\gamma p \to \pi^0\pi^0 p$ reaction.
 The MC event generation included the $\Delta \pi^0$ and $D_{13}(1520) \pi^0$
 intermediate states in double-$\pi^0$ photoproduction.}
\label{fig2}
\end{center}
\end{figure}
The details of the kinematic-fit parametrization
of the detector information and resolution were given in Ref.~\cite{slopemamic}.
The four- and five-cluster events that satisfied the hypothesis of the process
$\gamma p \to \pi^0\pi^0p \to 4\gamma p$ at the 2\% confidence level, CL,
(i.e., with a probability of misinterpretation
less than 2\%) were accepted as the reaction candidates.
The kinematic-fit output for which the pairing combination
of the four photons to two $\pi^0$s had the largest CL
was used to reconstruct the reaction kinematics.
Possible background from other reactions was checked by
their simulation and by determination of a probability for
them to be misidentified as $2\pi^0$ events.
Below the $\gamma p\to \pi^0\eta p$ threshold, the background contamination
was found to be only from interaction of incident photons in the windows of the
target cell and from random coincidences. This contamination was subtracted from
the experimental spectra by using data samples with
random coincidences and with an empty (no liquid hydrogen) target.
The background from the $\gamma p\to \pi^0\eta p\to 4\gamma p$ events
was estimated to be quite small, reaching only 0.5\% at incident-photon
energies of 1.4~GeV. So this background was neglected in our results.

The determination of the experimental acceptance was based on
a Monte Carlo (MC) simulation of the $\gamma p \to \pi^0\pi^0 p$ reaction
with different event generators based on various assumptions about the reaction
dynamics. For the most part, the MC  simulation was made as
the process $\gamma p \to \Delta \pi^0 \to \pi^0\pi^0 p$,
using the mass (1210~MeV) and width (100~MeV) of the Delta resonance at its pole
position (Ref.~\cite{PDG}). With these parameters, the agreement between the experimental 
and MC-simulation distributions of the invariant mass $m(\pi^0p)$ is much better
than that obtained when the Breit-Wigner parameters of $\Delta$ from Ref.~\cite{PDG}
are used. The same parameters also give a good description of the $\Delta$ peak
seen in the reaction $\pi^- p \to \pi^0\pi^0 n$~\cite{n2pi0}.
One simulation was made by generating an isotropic angular
distribution of the $\Delta\to \pi^0 p$ decay.
Another was made similar to the experimental angular distribution
in the region of the first peak in the $\gamma p \to \pi^0\pi^0 p$
total cross section (see Fig.\,\ref{fig1}).
Part of the MC simulation modelled the process
$\gamma p \to D_{13}(1520) \pi^0 \to \pi^0\pi^0 p$,
where the $D_{13}$ mass and width were taken as 1510~MeV and 110~MeV, respectively.
This simulation was used only in the analysis of the data taken
with the 1508-MeV electron beam and only for the energies in which
the $\gamma p \to D_{13}(1520) \pi^0 \to \pi^0\pi^0 p$ contribution
becomes visible. For these energies, the determination
of the experimental acceptance was done by mixing
the $\Delta$ and $D_{13}$ simulations, where the weights
of each simulation were adjusted to get the best
agreement with the experimental $m(\pi^0p)$ distributions.
The comparison of the measured and simulated $m^2(\pi^0p)$ distributions
is shown in Fig.\,\ref{fig2} for four different energies.
The agreement between the measured and simulated distributions
is better at lower energies.
This agreement is almost independent of the choice of the angular distribution used
for the$\Delta$-decay simulation; it affects only the shape of the $m(\pi^0\pi^0)$
spectrum. The resonance peaks seen in the $m^2(\pi^0p)$ spectra
look different from the Breght-Wigner shape as every event is represented by
two $m^2(\pi^0p)$ values, which are located symmetrically in the $\pi^0\pi^0p$ Dalitz plot
with respect to its symmetry line (see Ref.~\cite{n2pi0} and its Fig.~6 for more details).
Then every resonance band in the Dalits plot has its reflection with respect
this symmetry line. In Fig.\,\ref{fig2}(a) for example,
the projection of the $\Delta$ band to the $m^2(\pi^0p)$ axis is
seen in the right part of the spectrum, while
a bump in the left part just corresponds to the reflection of the $\Delta$ band
with respect to the symmetry line.

For the data at each electron-beam energy, the corresponding MC events
were propagated through a {\sc GEANT} (version 3.21) simulation of the experimental
setup, folded with resolutions of the detectors and conditions of the trigger.
The resulting simulated data were then analyzed in the same way as
the experimental data. The average acceptance for the data with
the 855-MeV electron beam was found to be close to 60\% for the entire
energy range of double-$\pi^0$ photoproduction.
The average acceptance for the data with the 1508-MeV electron beam
decreases smoothly from 55\% at $E_\gamma=617$~MeV to 42\%  at $E_\gamma=1400$~MeV.

The total cross sections obtained from the $\gamma p \to \pi^0\pi^0 p$
reaction are shown in Fig.\,\ref{fig1}
as a function of the incident-photon energy and are compared to some previous
measurements. The majority of previous $\gamma p \to \pi^0\pi^0 p$
experiments were at MAMI~\cite{Bragh,Hart,Wolf,Kot,Ahren2,Sara,Schum}.
In Fig.\,\ref{fig1}, we include only the most recent results
obtained with three different experimental setups: DAPHNE\,\cite{Ahren2}, TAPS\,\cite{Sara},
and Crystal Ball/TAPS\,\cite{Schum}. The other measurements were performed at
ELSA\,\cite{Thom,Sara} and by GRAAL\,\cite{Assa}. The results obtained in this work
are in good agreement with all previous measurements within the given statistical and 
systematic uncertainties. It was possible with our new data to reduce considerably the 
energy binning as well as the statistical uncertainties. The agreement of our 
total-cross-section results from the two measurements with different electron-beam 
energies can be seen in the overlapping range from $E_{\gamma}=617$~MeV to 
$E_{\gamma}=796$~MeV (shown later in Fig.\,\ref{fig7}).

The systematic uncertainties in the total and differential cross sections were estimated
to be not larger than 6\% and are dominated by the determination
of the experimental acceptance for $\gamma p \to \pi^0\pi^0 p$
and the photon-beam flux.
The systematic uncertainty because of the acceptance determination was studied by
comparing our results for the total cross sections that were obtained with
various MC simulations based on event generators with different $\gamma p \to \pi^0\pi^0 p$
dynamics. Also, we compared the total cross sections that were obtained from
the integration of the differential cross sections, which will be shown later in the text.
The systematic uncertainty in the photon-beam flux was determined
mostly by the variation of the tagging efficiency during the data-taking period.

\section{The Model}\label{formal}

\begin{figure}
\begin{center}
\resizebox{0.35\textwidth}{!}{%
\includegraphics{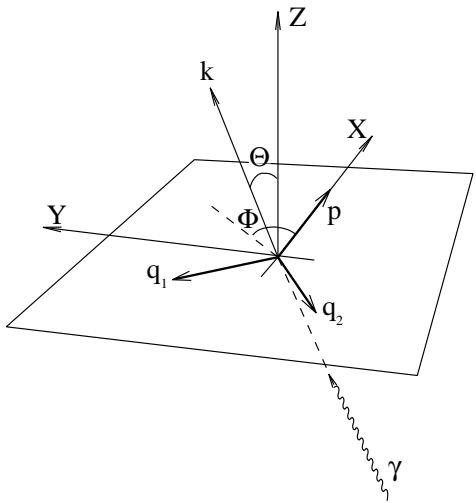}}
\caption{Definition of the coordinate system used in the present formalism.
 $\vec{k}$, $\vec{p}$, $\vec{q_1}$, and $\vec{q_2}$ are respectively
 three-momenta of the incident-photon, out-going proton, two pions in the 
 center of mass system. Axis $Z$ is a normal to the decay plane.  
 Axis $X$ is along $\vec{p}$. $\Theta$ and $\Phi$ are respectively the polar 
 and azimuthal angles of $\vec{k}$.}
\label{fig3}
\end{center}
\end{figure}

The formalism used to interpret our experimental data is described in
Ref.\,\cite{FiAr}, where the formal expressions are derived for the helicity amplitude as
well as for the cross section. At a particular photon energy, the reaction amplitude is
determined by four independent continuous variables for unpolarized experiments,
described below, and two discrete variables, which are taken as the initial and final
helicities of the nucleon. Our choice of coordinate system
is shown in Fig.\,\ref{fig3}, where all particles are in the center of mass frame.
Axis $Z$ is chosen along the
normal to the decay plane, which is defined by the three final-state particles. Axis $X$
is chosen along the outgoing-nucleon momentum. Angles $\Phi$ and $\Theta$
defined in  Fig.\,\ref{fig3} specify the direction of the incident-photon momentum
$\vec{k}$ in this coordinate system.
Together with the two angles, the energies of the two pions, $\omega_1$ and $\omega_2$,
uniquely determine the final-state kinematics. The angular dependence of the matrix
element is then given \cite{FiAr} by
\begin{eqnarray}\label{14}
T_{\nu\mu}(\omega_1,\omega_2;\Phi,\Theta)&=&\sum\limits_{JM}
t^{JM}_{\nu\mu}(\omega_1,\omega_2)\nonumber\\
&\times&D^J_{M\mu}(\Phi,\Theta,-\Phi)\,,
\end{eqnarray}
where $D_{m_1m_2}^j$ are the Wigner functions and $JM$ denote respectively
the total angular momentum and its projection on axis $Z$. The complex partial amplitudes
$t_{\nu\mu}^{JM}$, which depend on the energies $\omega_1$ and $\omega_2$,
contain the full dynamics of the process.

After spin summation and appropriate integration over $\omega_1$ and $\omega_2$,
one obtains the unpolarized differential cross section $d\sigma/(d\cos\Theta\,d\Phi)$
or the corresponding normalized quantity \cite{FiAr}
\begin{eqnarray}\label{WThetPhi}
&&W(\Theta,\Phi)\equiv\frac{1}{\sigma}\frac{d\sigma}{d\Omega}\nonumber\\
&&\phantom{xx}=\sum\limits_{L\geq 0}\sum\limits_{M=-L}^L\sqrt{\frac{2J+1}{4\pi}}\
W_{LM}Y_{LM}(\Theta,\Phi)\,,
\end{eqnarray}
which is expanded over spherical harmonics with $W_{00}=1$.
The coefficients $W_{LM}$ in Eq.\,(\ref{WThetPhi})
are hermitian combinations of the partial-wave amplitudes $t^{J M}_{\nu\mu}$. 
The corresponding expression was obtained in \cite{FiAr}:
\begin{eqnarray}
&&W_{LM}=\frac{\pi}{\sigma}{\cal K}\int d\omega_1d\omega_2\sum_{\nu\mu}
\sum_{JJ'M_JM_J'}(-1)^{M+\mu} \nonumber\\
&&\times C_{J'M_J'JM_J}^{LM}C_{J'\mu J-\mu}^{L0}t_{\nu\mu}^{J'M_J'}(\omega_1,\omega_2)^*
t_{\nu\mu}^{JM_J}(\omega_1,\omega_2)\,,\phantom{xxx}
\end{eqnarray}
where ${\cal K}$ is an appropriate phase space factor.  
Formula (\ref{WThetPhi}) determines the general structure of an angular distribution
in a manner analogous to the expansion of the cross section for single-meson 
photoproduction in terms of the Legendre polynomials.

To limit the number of model parameters, only the lowest partial waves
were used. Their choice is motivated by previous isobar-model analyses which demonstrated
that only waves with $J\leq 3/2$ were important below $E_\gamma=0.8$~GeV \cite{Oset,Ochi,FA}.

Expansion (\ref{WThetPhi}) written for the case of $J\leq 3/2$ is
\begin{eqnarray}\label{J32}
&&W(\Theta,\Phi)=\frac{1}{4\pi}\Bigg\{1-\frac{3}{\sqrt{2}}W_{11}P_1^1(\cos\Theta)
\cos\Phi\nonumber\\ &&+5\bigg(W_{20}P_2^0(\cos\Theta)
+\frac{1}{\sqrt{6}}W_{22}P_2^2(\cos\Theta)\cos 2\Phi\bigg)\nonumber \\
&&-\frac{7}{3}\bigg(\sqrt{3}W_{31}P_3^1(\cos\Theta)\cos\Phi\nonumber\\
&&-\frac{1}{2\sqrt{5}}W_{33}P_3^3(\cos\Theta)\cos 3\Phi\bigg)\Bigg\}\,,
\end{eqnarray}
from which it is easy to see that
the $\cos\Theta$ distribution has a general form
\begin{equation}\label{AB}
W(\cos\Theta)=A+B\cos^2\Theta\,,
\end{equation}
with
\begin{equation}
A=\frac12\left(1-\frac52 W_{20}\right)\,,\quad B=\frac{15}{4}W_{20}\,.
\end{equation}

\begin{figure}
\begin{center}
\resizebox{0.48\textwidth}{!}{%
\includegraphics{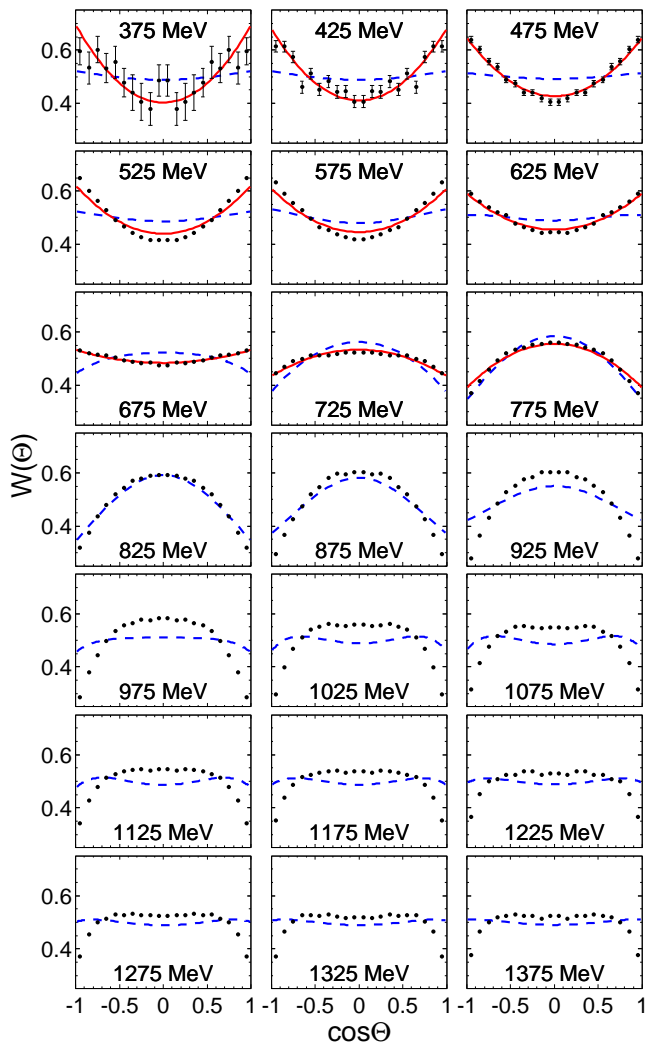}}
\caption{(Color online) Distribution $W(\Theta)=\int W(\Theta,\Phi)d\Phi$ shown
 as a function of $\cos\Theta$, where $\Theta$ is the polar angle of the incident
 photon in the coordinate frame presented in Fig.\,\protect\ref{fig3}. Our
 experimental results with statistical uncertainties are shown by filled circles.
 The predictions from the model of Ref.\,\protect\cite{FA} are shown by dashed lines.
 The results of fitting our data below $E_\gamma=0.8$~GeV are shown by solid lines.
 The energy label in each panel indicates the central photon energy for each bin.}
\label{fig4}
\end{center}
\end{figure}

\begin{figure}
\begin{center}
\resizebox{0.48\textwidth}{!}{%
\includegraphics{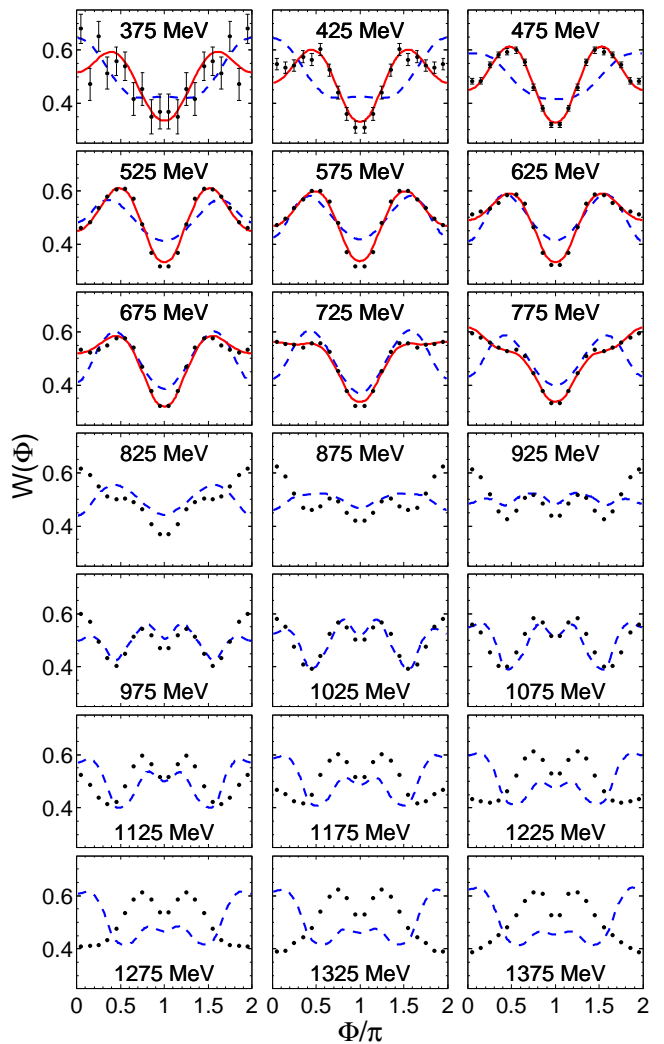}}
\caption{(Color online) Distribution $W(\Phi)=\pi\int W(\Theta,\Phi)\sin\Theta d\Theta$,
 where $\Phi$ is the azimuthal angle
 of the incident photon in the coordinate frame presented in Fig.\,\protect\ref{fig3}.
 Other notations are the same as in Fig.\,\protect\ref{fig4}.}
\label{fig5}
\end{center}
\end{figure}
As shown in Ref.\,\cite{FiAr}, the identity of the two pions together with parity
conservation results in the following symmetry relations:
\begin{equation}\label{Wsym}
W(\Theta,\Phi)=W(\pi-\Theta,\Phi)=W(\Theta,2\pi-\Phi)\,.
\end{equation}
Then, using the known properties of the spherical harmonics,
\begin{equation}
Y_{LM}(\pi-\Theta,\Phi)=(-1)^{L+M}Y_{LM}(\Theta,\Phi)\,,
\end{equation}
\begin{equation}
Y_{LM}(\Theta,2\pi-\Phi)=Y^*_{LM}(\Theta,\Phi)\,,
\end{equation}
one can see that the relations (\ref{Wsym}) lead to the following restrictions for the
coefficients $W_{LM}$
\begin{equation}\label{Wjm0}
W_{LM}=0\,,\ \mbox{if}\ L+M=\mbox{odd}\,,
\end{equation}
and
\begin{equation}\label{ImWjm0}
 Im(\ W_{LM})=0\,.
\end{equation}
In particular, $W_{L0}=0$ for $L=2n+1$. Furthermore, parity conservation requires that
the amplitudes with the same parity
interfere only in $W_{LM}$ with even $L$, whereas the waves
having the opposite parity interfere only in $W_{LM}$ with odd $L$.
This property was effectively used in partial wave analyses of inelastic pion-nucleon 
scattering $\pi N\to\pi\pi N$ \cite{Arnold,Morgan}.

The rule (\ref{Wjm0}) requires that, for example, the states with $J^P=\frac12^+$ produced
via $M1$ absorption (which in our case is saturated by the Roper resonance
and by the major part of the Born terms) can contribute only to $W_{00}$.
Therefore, in the region where states with $J\geq 3/2$ are not important, the angular
distribution $W(\Theta,\Phi)$ should be isotropic in both $\Theta$ and $\Phi$.

\section{Discussion of the results}

\begin{figure}
\begin{center}
\resizebox{0.48\textwidth}{!}{%
\includegraphics{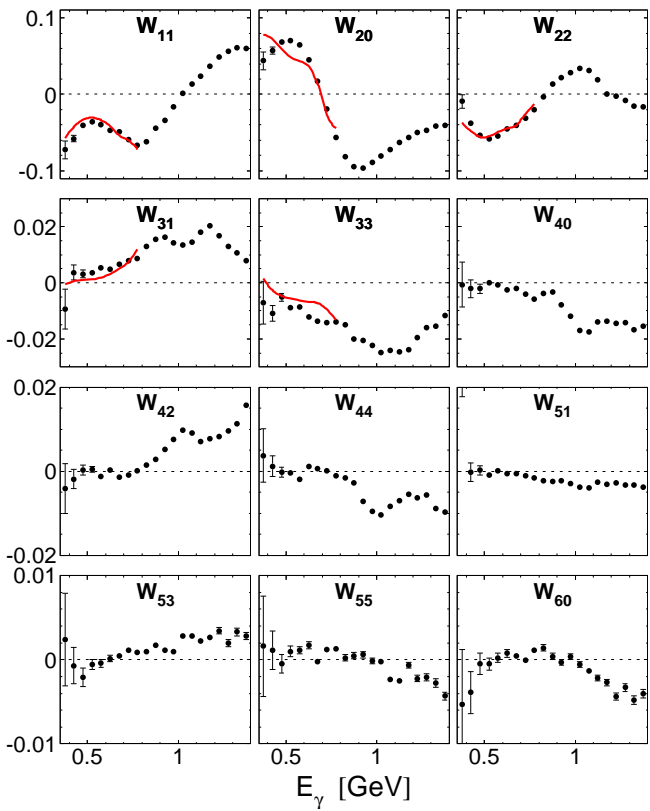}}
\caption{(Color online) Moments $W_{LM}$ (normalized such that $W_{00}=1$) as
 a function of the incident-photon energy. Our experimental results for the
 real part of $W_{LM}$ are shown by filled circles.
 The fit results are shown by solid lines.}
\label{fig6}
\end{center}
\end{figure}

\begin{figure}
\begin{center}
\resizebox{0.45\textwidth}{!}{%
\includegraphics{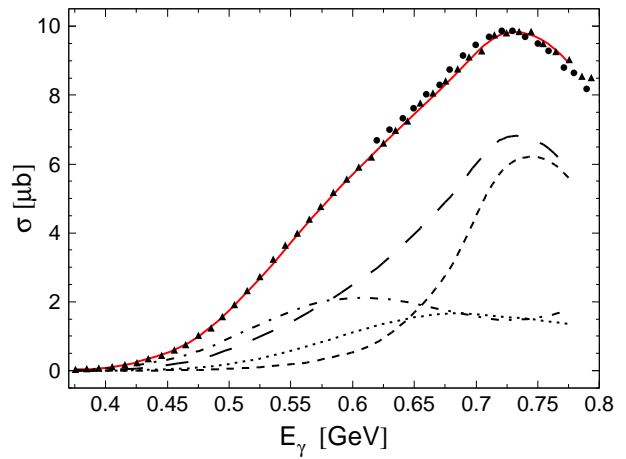}}
\caption{(Color online) Total cross section for $\gamma p\to\pi^0\pi^0 p$ as
 a function of the incident-photon energy.
 Our experimental results are shown by triangles and circles, respectively
 for the data with the 855-MeV and 1508-MeV electron beam.
 Only statistical uncertainties are shown.
 The fit results for the total cross section are shown by the solid line, and
 for the $3/2^-$, $3/2^+$, and $1/2^+$ waves by long-dashed, dash-dotted,
 and dotted lines, respectively. The $D_{13}(1520)$ contribution, calculated
 from the model of Ref.~\protect\cite{FA}, is shown by the short-dashed line.}
\label{fig7}
\end{center}
\end{figure}
The measured distributions of the angles $\Theta$ and $\Phi$ defined in
Sec.~\ref{formal}
are shown in Figs.\,\ref{fig4} and~\ref{fig5} for 50-MeV-wide energy bins.
The predictions that were made for these angular
distributions by the model from Ref.~\cite{FA} are shown in the same figures by dashed
lines. This model describes roughly the experimental distributions in the energy region
close to $D_{13}(1520)$. At the energies below $E_\gamma=900$~MeV, the shape of the
measured $\cos\Theta$ distributions shows good agreement with formula~(\ref{AB}).
However, at the energies below $D_{13}(1520)$, where the model of Ref.~\cite{FA} involves
the $P_{11}(1440)$ resonance and the nucleon Born terms come into play, the model
predicts an angular dependence that is weaker compared to the experimental data. As follows
from Eq.\,(\ref{AB}), the weakening of the moment $W_{20}$ leads to the model failure
at these energies. Within our approximation $J\leq3/2$, this moment is saturated by the
waves corresponding to the total angular momentum $J=3/2$. Therefore, this observation
indicates the persistence of such waves (and perhaps higher waves) at these energies. As
already discussed in Sec. I, the Roper resonance cannot
dominate at these energies. At the same time, the calculations from
Refs.\,\cite{Ochi,FA,Sara} predict a rapid fall of the $D_{13}(1520)$ contribution at
lower energies. This results in the significant underestimation of the measured total
cross section, which demonstrates almost linear energy dependence in this region.
According to the fit of Ref.\,\cite{Sara}, such behavior of the experimental data was
explained by a contribution from the $\Delta$-like resonance $D_{33}(1700)$, which
dominated the $\pi^0\pi^0$ channel in the full energy region considered.

Using Eq.\,(\ref{WThetPhi}), the expansion coefficients $W_{LM}$ were obtained from the
experimental two-dimensional plots of $\cos\Theta$ versus $\Phi$. To illustrate the
partial-wave content of the $\gamma p \to \pi^0\pi^0 p$ amplitude in more detail,
we show in Fig.\ref{fig6} the variation of these coefficients in the energy range
$E_\gamma=400-1400$~MeV for the waves with $J\leq 5/2$.
As remarked above, the values of the coefficients with odd $J$ are determined by the
interference of the states with different parities. If the insignificance of the waves
with $J>3/2$ is assumed, nonzero $W_{3M}$ coefficients arise from
the interference between $3/2^-$ and $3/2^+$. As one can see in
Fig.\,\ref{fig6}, $W_{31}$ and $W_{33}$ are quite small. This observation, for example, may
point to a predominantly background nature of the partial wave with $J^P=3/2^+$, which thus 
has a small imaginary part, whereas $J^P=3/2^-$ is mostly imaginary because of the closeness
to the $D_{13}(1520)$ pole. Furthermore, as will be shown later, the weakness of the
interference between $3/2^-$ and the positive-parity states $1/2^+$ and $3/2^+$ results
in a small forward-backward asymmetry in the angular distributions for
the final-state pions.

\begin{figure}
\begin{center}
\resizebox{0.48\textwidth}{!}{%
\includegraphics{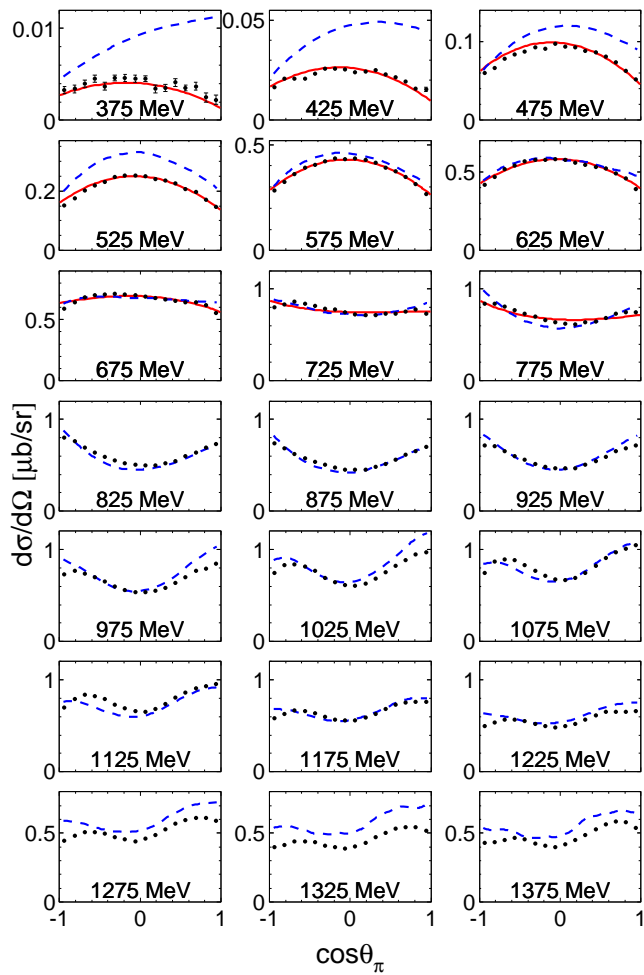}}
\caption{(Color online) $\gamma p\to\pi^0\pi^0 p$ differential cross sections as
 a function of the production angle of the outgoing $\pi^0$ in the center of mass frame.
 Since there are two identical pions, each cross section represents the average
 of two distributions. Our experimental results with statistical uncertainties are
 shown by filled circles. The predictions from our model are shown by solid lines.
 The dashed lines result from the Bonn-Gatchina model \cite{Sara,Thom}.}
\label{fig8}
\end{center}
\end{figure}

\begin{figure}
\begin{center}
\resizebox{0.48\textwidth}{!}{%
\includegraphics{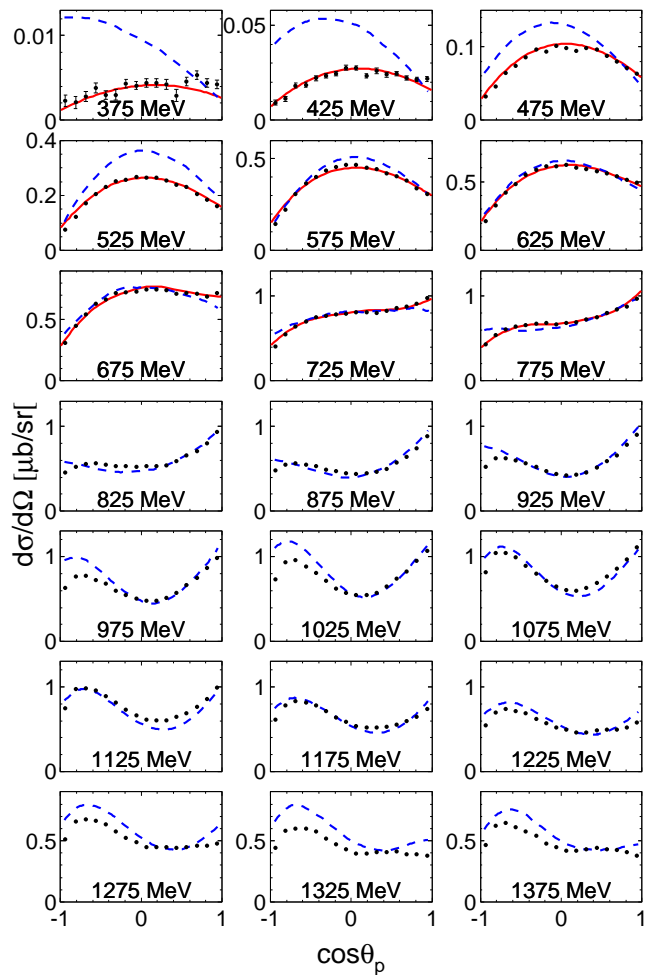}}
\caption{(Color online) Same as Fig.\,\protect\ref{fig8} but for the outgoing proton.}
\label{fig9}
\end{center}
\end{figure}

Already at low energies, the quantities $W_{20}$ and $W_{22}$, which (within our
restriction $J\leq 3/2$) are determined exclusively by the incoherent sum of the
states $3/2^-$ and $3/2^+$, achieve relatively large values. This observation indicates an
additional strong $3/2^-$ contribution, interfering with $D_{13}(1520)$, and/or a large
fraction of $3/2^+$. The latter can come, for example, from $\Delta$ decaying to $\pi\Delta$,
followed by $\Delta\to\pi N$. We cannot also exclude a strong $D_{33}(1700)$
amplitude, as was found in Ref.\,\cite{Sara}. However, as discussed in
Sect. I, the experimental data on $\pi^+\pi^-$ photoproduction seem to leave no room
for strong coupling to this resonance.

The coefficient $W_{11}$, coming from the interference of the $3/2^-$ wave with the
positive-parity waves $1/2^+$ and $3/2^+$, demonstrates quite sharp energy dependence in
the region $E_{\gamma}=500-650$~MeV. The moments with $L=5$ are small.

\begin{figure}
\begin{center}
\resizebox{0.48\textwidth}{!}{%
\includegraphics{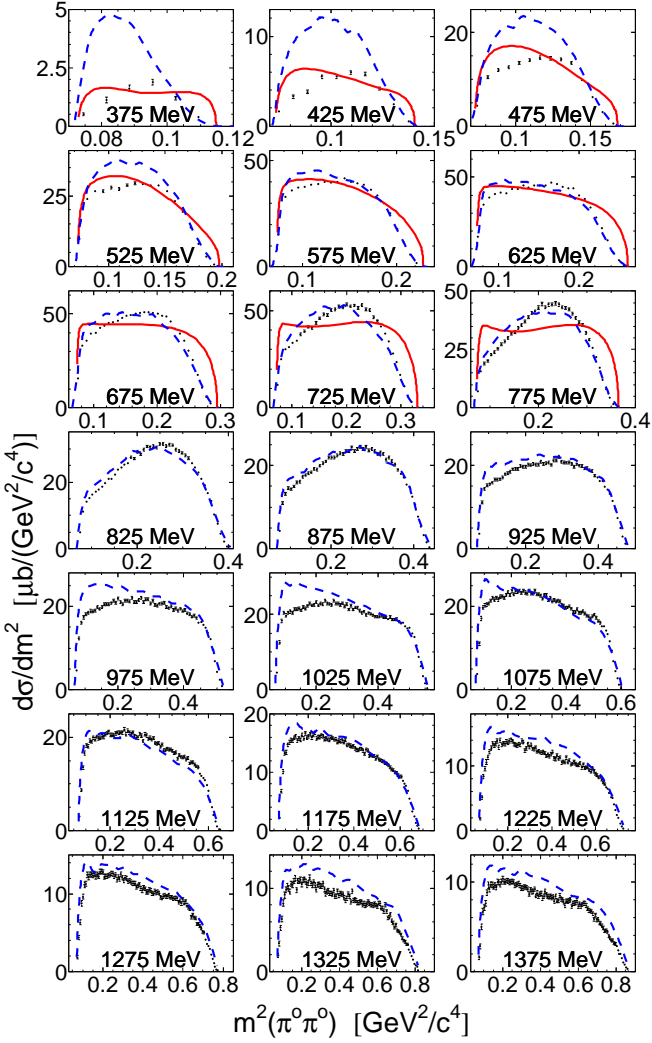}}
\caption{(Color online) $\gamma p\to\pi^0\pi^0 p$ differential cross sections as
 a function the invariant mass squared $m^2(\pi^0\pi^0).$
 Our experimental results with statistical uncertainties are shown by filled circles.
 The predictions from our model are shown by solid lines.
The dashed lines result from the Bonn-Gatchina model \cite{Sara,Thom}.}
\label{fig10}
\end{center}
\end{figure}
\begin{figure}
\begin{center}
\resizebox{0.48\textwidth}{!}{%
\includegraphics{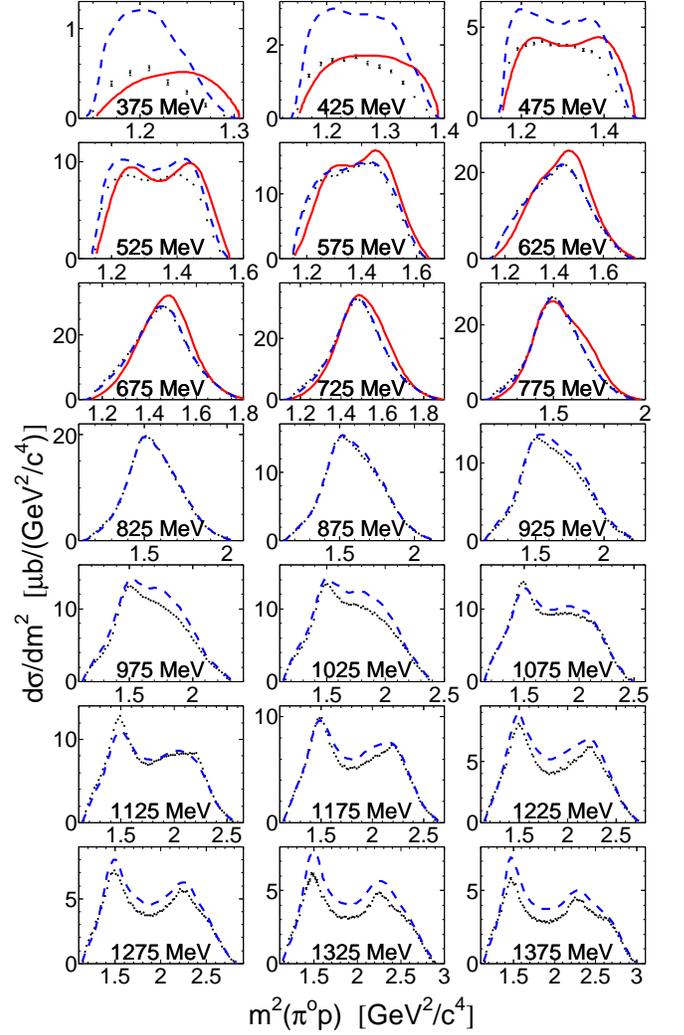}}
\caption{(Color online) Same as Fig.\,\protect\ref{fig10}, but for $m^2(\pi^0 p)$.
 Since there are two identical pions, each cross section represents the average
 of two distributions.}
\label{fig11}
\end{center}
\end{figure}
\begin{figure}
\begin{center}
\resizebox{0.48\textwidth}{!}{%
\includegraphics{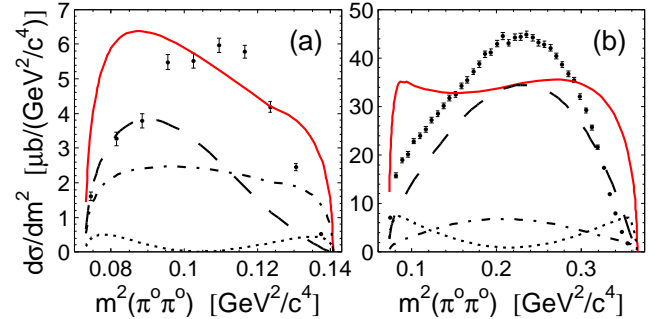}}
\caption{(Color online) The partial wave contributions to the $\pi\pi$ spectrum
 for $E_\gamma = 425$ (a) and $775$ MeV (b) (enlarged from Fig.~\protect\ref{fig10}).
 Long-dashed, dash-dotted, and dotted lines correspond to the $3/2^-$, $3/2^+$,
 and $1/2^+$ waves, respectively. The sum of all waves is shown by solid lines.}
\label{fig12}
\end{center}
\end{figure}

To fit the measured values of the moments $W_{LM}$, we assumed the model in which the final
$\pi\pi N$ state is produced exclusively via the intermediate $\pi\Delta$ state. First,
the resonance $D_{13}(1520)$, whose role in this reaction is more or less firmly
established, was put into the $3/2^-$ wave. The corresponding amplitude was parametrized
in the Breit-Wigner form, with parameters taken from PDG\,\cite{PDG}.
The only other partial waves included were those that lead to $s$ and $p$ wave in the final
$\pi\Delta$ state: $J^P=1/2^+$, $3/2^-$, and $3/2^+$. From our fit, the wave $J^P=5/2^+$, 
containing
$\pi\Delta$ in a $p$ state, is negligibly small and was
excluded from further consideration. Each partial-wave amplitude was parametrized in the form
\begin{equation}\label{tJM}
t_{J^P}=\Big[t_B+t_R(W)\Big]\,G_\Delta F_{\Delta\to\pi N}\,,
\end{equation}
where the two terms in the brackets stand for a smooth background and a rapidly varying
part $t_R(W)$, which can contain $s$-channel resonances. The factors $G_\Delta$ and
$F_{\Delta\to\pi N}$ are respectively the propagator and the $\pi N$-decay vertex of the
$\Delta$ isobar. The energy-independent background in each partial wave was parametrized as
\begin{equation}
t_B=\left|t_B\right|e^{i\phi_B}\,,
\end{equation}
with adjustable constants $\left|t_B\right|$ and $\phi_B$, whereas
 the rapidly-varying part was taken as
\begin{equation}
t_R(W)=\left|t_R(W)\right|e^{i\phi_R(W)}
\end{equation}
with
\begin{equation}\label{gPhi}
\left|t_R(W)\right|=a_0+a_1q+a_2q^2\,,\quad \phi_R(W)=\frac{q^3}{b_1+b_2q^2}\,,
\end{equation}
where $q$ is the maximum pion momentum in the final $\pi\pi N$ state,
corresponding to the total energy $W$ in the center of mass system
\begin{equation}
q=\frac{\sqrt{\Big(W^2-(M_N+2m_\pi)^2\Big)\Big(W^2-M_N^2\Big)}}{2W}\,.
\end{equation}
The coefficients $a_i$ and $b_i$ in Eq.\,(\ref{gPhi}) should be determined from the fit.
Since our fit was restricted to a limited energy range, the parametrization of
$\left|t_R(W)\right|$ by a simple polynomial formula (\ref{gPhi}) was expected to be
satisfactory.

The results of the fit to the $W_{LM}$ moments are shown by solid lines in
Fig.\,\ref{fig6}. Instead of listing the results for the fit parameters, the
integrated partial cross sections $\sigma_{J^P}$ are shown in Fig.\,\ref{fig7}.
As expected, the strongest $\pi^0\pi^0$ production into the entire energy region
comes from $J^P=3/2^-$ and $3/2^+$ (shown by the long-dashed and dash-dotted lines in
Fig.\,\ref{fig7}). At lower photon energy energies
the partial cross section $\sigma_{3/2^-}$ falls off slower
than predicted by the model \cite{FA}.
In addition, the wave $3/2^+$ turns out to be very important,
especially at the energies below $E_\gamma=650$ MeV.

\begin{figure}
\begin{center}
\resizebox{0.45\textwidth}{!}{%
\includegraphics{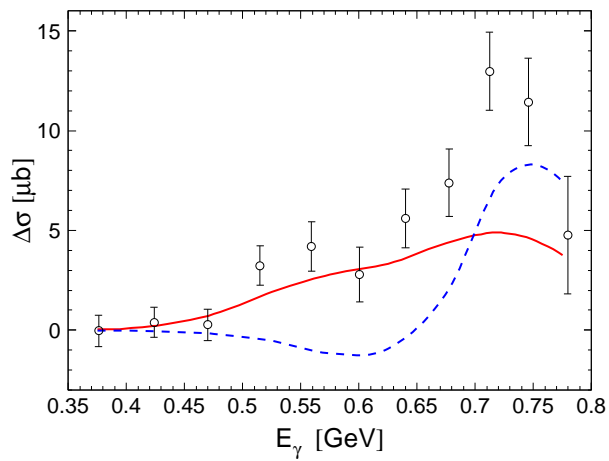}}
\caption{(Color online) Helicity asymmetry $\Delta\sigma=\sigma_{3/2}-\sigma_{1/2}$
 for $\gamma p\to\pi^0\pi^0 p$.
 The experimental data from Ref.\,\protect\cite{Ahren2} are shown by open circles.
 The prediction is shown by the solid line for our model, and
 by the dashed line for the isobar model of Ref.\,\protect\cite{FA}.}
\label{fig13}
\end{center}
\end{figure}

After fitting the model parameters to the measured total cross sections and moments
$W_{LM}$, the reliability of our parametrization [given by
Eqs.~(\ref{tJM})--(\ref{gPhi})] was checked by comparing the model predictions with the
$\gamma p\to\pi^0\pi^0 p$ experimental results for other observables. These comparisons
are shown in Figs.~\ref{fig8}--\ref{fig13}. Before discussing the agreement between
the experimental data and the calculation, it is important to note that the values of
$W_{LM}$ do not determine final-state distributions of $\gamma p\to\pi^0\pi^0 p$ [because
of the integration over energies $\omega_1$ and $\omega_2$ in Eq.(\ref{WThetPhi})].
Therefore, the theoretical results shown in Figs.~\ref{fig8}--\ref{fig13}
depend essentially on the model used for describing the production mechanism. As discussed
above, we used the assumption that the $\gamma p\to\pi^0\pi^0 p$ reaction proceeds exclusively
through the transition $\Delta\pi\to\pi\pi N$.
Agreement with the measurements would support this asumption.

In Figs.\,\ref{fig8} and \,\ref{fig9}, we show our experimental results
for the $\gamma p\to\pi^0\pi^0 p$ differential cross sections as a function
of the production angle of the outgoing $\pi^0$ and proton in the center of mass frame.
Figures\,\ref{fig10} and \,\ref{fig11} show our
differential cross sections as a function of
the invariant mass squared $m^2(\pi^0\pi^0)$ and $m^2(\pi^0p)$.
These results are obtained for the same energies
that were used for the angular distribution, shown in Figs.\,\ref{fig4} and~\ref{fig5}.
The model predictions are shown in these figures up to $E_\gamma=775$~MeV.
They are in reasonable agreement with the experimental results, especially concerning 
the shape of the angular distributions.
In particular, the calculations reproduce not only the convexity and its sign, which changes
with energy, but also the forward-backward asymmetry. This asymmetry is mostly determined by
the interference of the $3/2^-$ wave with positive-parity waves
(in our case $1/2^+$ and $3/2^+$). As discussed above, the smallness of this forward-backward
asymmetry indicates the weakness of this interference. This is also related to the small
values of the moments $W_{3M}$, shown in Fig.\,\ref{fig6}.

The model predictions for the invariant-mass distributions,
shown in Figs.\,\ref{fig10} and \,\ref{fig11}, are not so impressive.
The poorer level of agreement could be partially explained by
pion rescattering in the final state, which
was neglected in our model. According to Refs.\,\cite{Bernard,Roca1},
the pion loops in the $\pi^0\pi^0$ channel can lead
to a significant enhancement of the cross section at low energies. This is primarily
because of a large yield of $\pi^+\pi^-$ pairs, which in turn can rescatter into neutral
pions. As known, the interaction between pions in the state $J^P=0^+,I=0$ is attractive.
The corresponding phase shift reaches $\pi/2$ close to $M_{\pi\pi}=900$~MeV
(see Ref.\,\cite{Protopop}).
Therefore, it is reasonable to expect that the inclusion of this effect will shift
the $m(\pi^0\pi^0)$ spectrum to higher masses.
Another possible reason for the poor agreement is that the fraction of the wave $J^P=1/2^+$
in our model is slightly overestimated. If the $\pi\pi$ system
does not resonate (or the $\pi\pi$ resonance is wide, like $f_0(600)$),
then the shape of the Dalitz plot $(M_{\pi\pi}^2,M_{\pi p}^2)$ is totally
determined by the spin-parity $J^P$ of a given partial wave (see
the corresponding discussion for $\pi^0\eta$ photoproduction in
Ref.\,\cite{FOT}). To illustrate this statement, the contributions of the individual states
to the $\pi\pi$ spectrum are shown in Fig.\,\ref{fig12} for $E_\gamma=425$ and 775 MeV.
The predicted enhancement of the cross section at the boundaries of
the kinematical region, which is typical for the contribution
from the state $J^P=1/2^+$, is not exhibited by the experimental data.
In this respect, our experimental results prefer a $1/2^+$ fraction
that is even smaller than predicted by the fit of the moments $W_{LM}$.

In Figs. \ref{fig8} to \ref{fig11}, our experimental results are also
compared to the predictions of the Bonn-Gatchina model \cite{Sara,Thom}.
The approach of Refs.~\cite{Sara,Thom} is based on the event-by-event likelihood
fit that allows one to take accurately into account the correlations between the
different reaction channels, for example, $\pi\Delta$ and $\sigma N$. As one can
see, the model from Refs.~\cite{Sara,Thom} describes our experimental results
quite well above 550 MeV. Only at low $E_\gamma$ does it overestimate
the measured cross sections.

In Fig.\,\ref{fig13}, the prediction of our model for the
$\Delta\sigma=\sigma_{3/2}-\sigma_{1/2}$ helicity asymmetry
is compared to the $\gamma p\to\pi^0\pi^0 p$ experimental data from Ref.\,\cite{Ahren2}.
The experimental data, measured as the difference between the total cross sections with the
initial $\gamma p$-system helicity $3/2$ and $1/2$,
indicate the dominance of the $\lambda=3/2$ component over $\lambda=1/2$ in the
energy region $W=1400-1500$~MeV, excluding any large contribution from the $J=1/2$
waves. Our model reproduces the general trend of the data, which shows $\sigma_{3/2}$
dominance.
The isobar model of Ref.~\cite{FA}, in which the Roper
resonance was rather important in the region $E_\gamma=500-600$~MeV,
predicts negative values for $\Delta\sigma$ (shown by the dashed line in the same figure).
In this energy region, this is in contradiction with the experimental data.

\section{Summary and conclusions}\label{conclusion}

The $\gamma p\to\pi^0\pi^0 p$ reaction
has been measured at the tagged-photon facility
of the Mainz Microtron MAMI-C using the Crystal Ball
and TAPS spectrometers.
The experimental results, obtained from the production threshold
up to a photon energy of 1.4 GeV,
include the total cross sections, various differential
cross sections, and specific angular distributions.
The moments $W_{LM}$ obtained from these angular distributions
were used to study the importance of different partial waves in double-$\pi^0$ photoproduction
at energies below $D_{13}(1520)$ (a region that has not been so far fully understood
theoretically). The reliability of our model was checked by the comparison of its predictions
with the $\gamma p\to\pi^0\pi^0 p$ experimental results for other observables.

Our analysis of the energy dependence of $W_{LM}$ showed that a large contribution from the
$J=3/2$ waves is necessary not only in the region of $D_{13}(1520)$ but also at energies
below. According to our results, these waves seem to be responsible for an almost linear rise
of the $\gamma p\to\pi^0\pi^0 p$ total cross section in the region $E_\gamma=450-725$~MeV.
Isobar models with the dominant contribution from $D_{13}(1520)$ and a moderate role for
the Roper resonance cannot explain such features in double-$\pi^0$ photoproduction.
Whether these features are the reflection of a large $J^{\pi}=3/2^+$ fraction of
$\pi^+\pi^-\to\pi^0\pi^0$ rescattering,
or are a consequence of the strong $D_{33}(1700)$ excitation, found in Ref.\,\cite{Sara},
requires further experimental and theoretical studies.

\section*{Acknowledgment}

The authors wish to acknowledge the excellent support of the accelerator group and
operators of MAMI. This work was supported by the Deutsche Forschungsgemeinschaft (SFB
443, SFB/TR16), DFG-RFBR (Grant No. 09-02-91330), the European Community-Research
Infrastructure Activity under the FP6 ``Structuring the European Research Area''
program (Hadron Physics, Contract No. RII3-CT-2004-506078), Schweizerischer
Nationalfonds, the UK Sciences and Technology Facilities Council, U.S. DOE, U.S. NSF, and
NSERC (Canada). A.F. acknowledges additional support from the RF Federal programm
``Kadry''(Contract No. P691) and the MSE Program ``Nauka'' (Contract No. 1.604.2011).
We thank the undergraduate students of Mount Allison University
and The George Washington University for their assistance.


\end{document}